\documentclass[twocolumn]{emulateapj}
\usepackage{apjfonts}
\usepackage{hhline}
\usepackage{url}
\usepackage{amssymb}
\usepackage{soul}
\usepackage{xcolor}
 
\shorttitle{H$\alpha$ emitters in NGC 6397}
\shortauthors{Pallanca et al.}

\begin{document}

\title{A complete census of H$\alpha$ emitters  in NGC 6397}



\author{Cristina Pallanca\altaffilmark{1,2},
Giacomo Beccari\altaffilmark{3},
Francesco R. Ferraro\altaffilmark{1,2},
Luca Pasquini\altaffilmark{3},
Barbara Lanzoni\altaffilmark{1,2},
Alessio Mucciarelli\altaffilmark{1,2}}


\altaffiltext{1}{Dipartimento di Fisica e Astronomia, Alma Mater Studiorum Universit\`a di Bologna, Via Gobetti 93/2, I-40129 Bologna, Italy}
\altaffiltext{2}{INAF-Osservatorio Astronomico di Bologna, Via Gobetti 93/3, I-40129 Bologna, Italy}
\altaffiltext{3}{European Southern Observatory, Karl-Schwarzschild-Strasse 2, 85748 Garching bei M\"unchen, Germany}

\def\comA{COM J1740$-$5340A}

\begin{abstract}
We used a dataset of archival  Hubble Space Telescope images obtained through the F555W, F814W and F656N filters, to perform a complete search for objects showing $H\alpha$ emission in the globular cluster NGC 6397. As photometric diagnostic, we used the $(V-H\alpha)_0$ color excess in the  $(V-H\alpha)_0$-$(V-I)_0$ color-color diagram.
 In the analysed field of view, we identified 53 $H\alpha$ emitters.   In particular, we confirmed the optical counterpart to 20 X-ray sources (7 cataclysmic variables, 2 millisecond pulsars and 11 active binaries) and identified 33 previously unknown sources, thus significantly enlarging the population of known active binaries in this cluster.  We report the main characteristics for each class of objects.   Photometric estimates of the equivalent width of the  $H\alpha$ emission line, were derived from the $(V-H\alpha)_0$-excess  and, for the first time,  compared to the spectroscopic measurements obtained from the analysis of MUSE spectra. The very good agreement between the spectroscopic and photometric measures fully confirmed the reliability of the proposed approach to measure the $H\alpha$ emission. The search demonstrated the efficiency of this novel approach to pinpoint and measure $H\alpha$-emitters, thus offering a powerful tool to conduct complete census of objects whose formation and evolution can be strongly affected by dynamical interactions in star clusters.
\end{abstract}


\keywords{globular clusters: general --- globular clusters:
  individual (NGC 6397)}



\section{Introduction}
The crowded cores of globular clusters (GCs) are very efficient ``furnaces'' for generating exotic objects, such as cataclysmic variables (CVs), low-mass X-ray binaries (LMXBs),  millisecond pulsars (MSPs) and blue stragglers (see \citealt{bailyn95}). 
Most of these objects are thought to be the result of the evolution of various kinds of binary systems originated and/or hardened by stellar interactions \citep{davies05,ivanova06}. 
The nature and even the existence of binary by-products are strongly related to the cluster core dynamics (\citealt{ferraro03,ferraro09,ferraro15a}). 
Hence a complete census of these objects (whose formation and evolution are strongly affected by dynamical interactions) is crucial in order to trace the dynamical evolution of the parent cluster \citep[][]{beccari06,ferraro12,lanzoni16}

Most of  these exotic objects are expected to be, at least, potential $H\alpha$ emitters.
LMXBs are binary systems made of a compact object (a stellar black hole or a neutron star; NS)  accreting mass from a low mass companion through an accretion disk.  
They are usually bright X-ray sources and, in the case of a NS compact object, they are thought to be the progenitors of MSPs.
MSPs form in binary systems containing a NS first evolved through the LMXB phase and then eventually spun up  to millisecond periods by  mass accretion from the evolving companion (see e.g. \citealt{ferraro15b}), that, in turn, is expected to become a white dwarf ~\citep[e.g][]{lyne87, alpar82, bhattvan91}. 
However, in the last years the number of detections of non-degenerate companions significantly increased \citep{pallancaM28H, pallancaM28I, pallancaM5C,cadelanoM71A}.   Note that the first non-degenerate companion to a MSP in a GC (PSR J1740$-$5340A) was found by \citet{ferraro01com6397} in NGC 6397 (the cluster subject of this paper).
 
MSPs usually do not have strong $H\alpha$ excess, since accretion is inhibited by the presence of a strong magnetic field. 
However, in the cases of transitional-MSPs (T-MSPs), which are thought to be the evolutionary link between LMXBs and MSPs, a significant $H\alpha$ emission is not unexpected  and it has been indeed confirmed in  the case of  PSR J1824$-$2452I in the GC M28 \citep{pallancaM28I} as well as in the Redback system PSR J1740$-$5340A in NGC6397 \citep{ferraro01com6397,sabbi03}.
 
CVs are binary systems in which a white dwarf  is accreting material from a main sequence (MS) companion \citep[see][for a review]{knigge12}. 
From the theoretical point of view, models predict that  there should be 60$-$180 CVs for every $10^6 {\rm L}_\odot$ in an old stellar population \citep{townsley05}, most of which should be   formed dynamically~\citep{ivanova06}.
Several methods have been proposed to detect candidate CVs \citep{knigge12}:  (i)  studying the variability associated to dwarf nova outbursts; (ii) looking for objects showing bluer colors than MS stars, because of the energy  released in a hot region close to the white dwarf; (iii) searching for the X-ray emission produced by the accretion onto the compact object; (iv)  spectroscopically detecting emission lines. 
However, none of these methods is able to recover the entire CV population expected and in most cases  more than one method is needed to  properly classify the objects. 
\citet{beccari13} showed that the detection of candidate CVs as sources showing $H\alpha$ excess emission in GCs can be improved by using a proper combination of broad bands $V$ and $I$ and narrow band $H\alpha$ imaging.
This method was successfully applied in star-forming regions (see \citealt{demarchi10}) and in other clusters (see \citealt{pallancaM28I}) to identify  $H\alpha$ emitters.

It is worth mentioning that, in  addition to the classes of interacting binaries described above, in which the $H\alpha$ excess is  related to the mass transfer process, there are other objects which are expected to show $H\alpha$ emission, because of their intense chromospheric activity:  the so-called active binaries (ABs). The two most relevant prototypes of this category of objects  are  RS Canum Venaticorum and BY Draconis  (BY Dra) binaries. Indeed a large population of BY Dra-type binaries has been found in NGC6397 (see \citealt{taylor01,cohn10}).
 
In this paper we applied the $(V-H\alpha)_0-(V-I)_0$ color-color  method  to identify $H\alpha$ emitters and to derive a photometric estimate of the $H\alpha$ Equivalent Width (EW).

\section{The target cluster: NGC6397}

In  the framework of the study of exotic populations in dense stellar systems, NGC 6397 is an ideal GC to be studied,  because its X-ray population is already  known and classified.
Thanks to  deep Chandra imaging of the cluster, \citet{grindlay01} detected 25 X-ray sources with $L{\rm x}> 3 \times 10^{29}$ erg s$^{-1}$ located within 2\arcmin\ from the cluster center. 
In a more recent work \citet{bogdanov10} identified 79 Chandra X-ray sources that lie within the half-mass radius \citep[$r_h=2\arcmin.33$][]{harris96}, of which, 15 are CVs, 42 are ABs, 1 is a radio detected MSP and 1 is a candidate MSP \citep{cohn10}.
The comparison between the MSP and CV populations of NGC 6397 and 47 Tuc shows that the CV to MSP ratio is $\sim10$ times larger in NGC 6397, thus suggesting  a dichotomy of compact objects and binary production in the two clusters: NGC 6397 has overproduced CVs (for its mass), perhaps because of two-body captures during core collapse, while its relatively lower MSP production might be due to a lower NS retention or original fraction due to differences in the cluster initial mass functions, in turn due to their very different metallicities \citep{grindlay01}.

The CV population of NGC 6397 has been identified with different methods:
CVs 1-3 were originally identified  in a HST $H\alpha$ survey \citep{cool95}, 
CVs 4-5 were found via optical variability or as counterparts to ROSAT HRI sources \citep{cool98, grindlay99},
CVs 6-8 were discovered in a deeper follow-up HST $H\alpha$ survey \citep{grindlay01}, 
CV9 has been identified  on the basis of its Chandra spectrum, 
and the remaining CVs have been detected as Chandra X-sources \citep{bogdanov10}.

A spectroscopic confirmation through hydrogen and helium lines has been obtained for four objects   \citep[CVs 1-4;][]{grindlay95, edmonds99}. 
All these CVs show $He II \lambda4686$, a line seen almost exclusively in magnetic CVs and nova-like variables. 
For CV2 and CV3 the authors also observed nova-like eruptions, which are unexpected for magnetic CVs, since the magnetic field should prevent the disk instability that leads to the nova phenomenon.

Variability has been detected in CVs $1-9$, and in two cases it has been found to be particularly strong: CV2 shows a variability of at least 2.7 mag and CV3 is seen 1.8 mag brighter in eruption than in quiescence \citep{shara05}. 
On the other hand CV1 and CV6 were identified as eclipsing CVs by \citet{kaluzny03}. The light curves of both of them exhibit two distinct minima per orbital period and they seem to be dominated by ellipsoidal variations. 
The accretion rate in CV6 has been roughly constant over the observing seasons 2002-2004 \citep{kaluzny06}.

\citet{cohn10} showed the location of the CV candidates in the color-magnitude diagrams (CMDs) and suggested that there is an evolutionary sequence from young, bright CVs, to old, faint ones. 
The optical emission of the six brightest CVs appears to be dominated by a relatively massive secondary, while that of the faint CVs is mainly due to the white dwarfs, with very little contribution from a very low-mass secondary. CVs 1$-$6 are the young, recently formed systems. 

\citet{cohn10} 
reported on the detection of 42 candidate ABs, of which 25 lie within the field covered with the data used in this work.
Most of them  draw   a relatively homogenous binary sequence alongside the MS \citep{cohn10} and according to \citet{taylor01} they are likely BY Dra objects.

In this paper we applied the selection method of $H\alpha$ emitters commonly used in star forming regions \citep{demarchi10} and recently applied to GCs  \citep{beccari13,pallancaM28I} as a tool to both identify exotic objects  and  derive a photometric estimate of the EW (pEW) of the $H\alpha$ emission.

\section{Observations and data reduction}

The data used in this work (GO 7335 PI Grindlay) consist of 88 images obtained with the Wide Field Planetary Camera 2 (WFPC2) in the broad F555W and F814W bands and the narrow F656N filter (hereafter $V$, $I$ and $H\alpha$ respectively).

We analysed the entire set of images through a standard point spread function (PSF) fitting procedure by using DAOPHOT/ALLFRAME \citep{st87}. 
In short, we calculated a PSF model on each image using approximately 50 isolated and not saturated stars.
We then used the DAOPHOT/MONTAGE2 routine to stack all the $V$ and $I$ images together. In this way we obtained 4 very deep and very high signal-to-noise ratio images (one for each CCD of the WFPC2 detector) cleaned of cosmic rays and detector blemishes.
We extracted a list of stellar sources from these reference images and we used it as input master list of ALLFRAME~\citep{st94}, which simultaneously determines the star brightness in all frames. 
As last step, the magnitude of each star in each band was obtained as the average of at least 4 measures, while the standard deviation of the repeated measures was taken as photometric error.
We calibrated the WFPC2 magnitudes into the VEGAMAG system using the standard recipe and zeropoints reported in~\citet[][]{ho95}.
Finally, we de-reddened all observed magnitudes by assuming a color excess E$(B-V)=0.18$ \citep {gratton03,richer08}.
 
\begin{figure*}
\includegraphics[width=150mm]{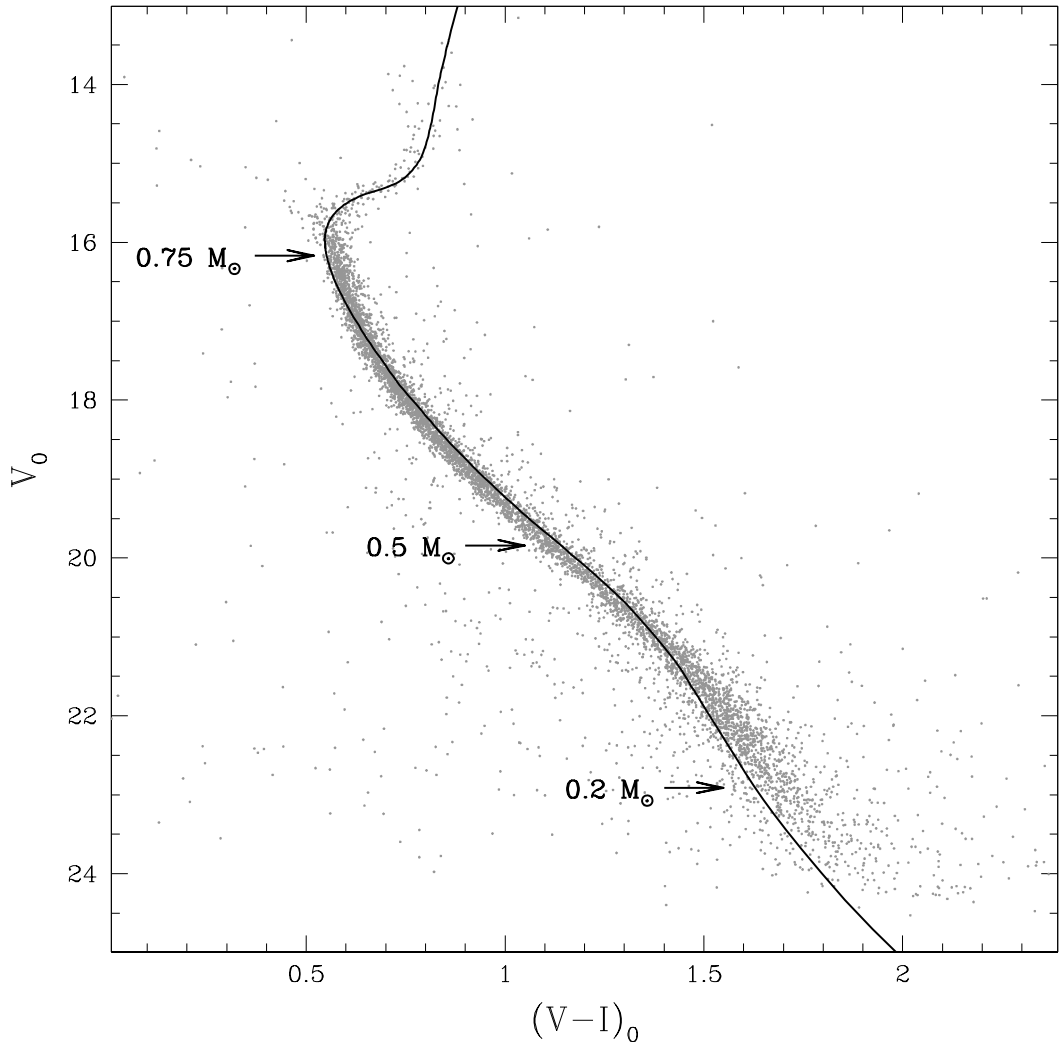}
 \caption{Reddening-corrected, optical color-magnitude diagram of NGC 6397.  
The black line indicates the location of the best-fit isochrone (age=13.5 Gyr, $[Fe/H]=-2.03$, $[\alpha/Fe]=0.4$; \citealp{richer08}) assuming a reddening of ${E(B-V)}=0.18$ and a distance modulus ${\rm (m-M)}_0=12.01$.  The location of stars with 0.2, 0.5 and 0.75M$_{\odot}$ masses are labelled and marked with the arrows.}
\label{cmdiso}
\end{figure*}

We show in Figure \ref{cmdiso} the  reddening-corrected  $V_0$ vs. $(V-I)_0$ CMD obtained from the WFPC2 photometric catalog (gray points).  
The cluster's stars are sampled  from the bottom  of the red giant branch down to $\sim$7 magnitudes below the MS Turn off ($V_0\sim16.2$).
The best-fit isochrone \citep{dotter08}, obtained for an age=13.5 Gyr, a metallicity [Fe/H]=$-$2.03 and [$\alpha$/Fe]=0.4 \citep{richer08}, and assuming a distance modulus ${\rm (m-M)}_0=12.01$ \citep{gratton03} is also shown in Figure \ref{cmdiso} (solid line). 
The isochrone has been used to convert the sampled magnitudes into stellar masses. 
As shown in Figure \ref{cmdiso} we are able to sample the MS stars from 0.75  M$_\odot$ (corresponding to the mass of stars at the Turn off) down to 0.2 M$_\odot$ at the bottom of the MS. 

We finally transformed the relative stellar positions into absolute Right Ascension (RA) and Declination (Dec) using more than 300 stars in common with a photometric catalog  \citep{ferraro01com6397}   from ground-based images obtained with the Wide Field Imager (WFI) at the 2.2m telescope at La Silla. 
This allowed us to obtain a global astrometric accuracy of ~0\farcs3 in both RA and Dec. 
It is worth to notice that the same astrometric accuracy allowed~\citet[][]{ferraro01com6397} to identify the optical counterpart of the MSP J1740$-$5340  in the same HST WFPC2 dataset used in this paper.
 
\section{Selection of  H$\alpha$ emitters and equivalent width measure}

\subsection{Photometric approach}\label{sec:photo}
\begin{figure*}
\includegraphics[width=150mm]{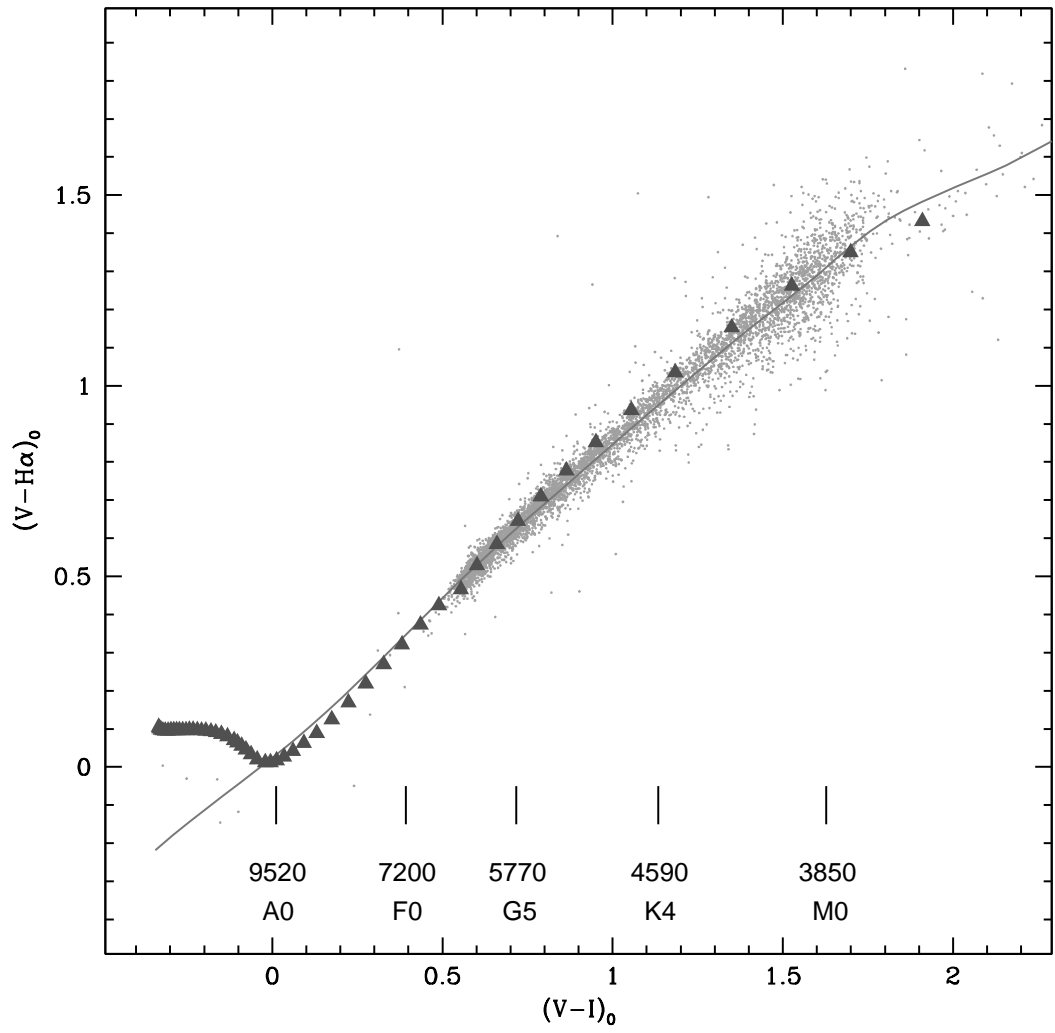}
\caption{Reddening-corrected color-color diagram.  
The solid line is the median color of stars with no $H\alpha$ excess (gray dots) and hence the location of stars with EW$H\alpha$ = 0.  
It corresponds well to the location (gray triangles) predicted by atmospheric models \citep{bessell98}.  
On the bottom part of the plot the MS spectral types and effective temperatures (in Kelvin) are reported for reference.}
\label{colcol}
\end{figure*}

\begin{figure*}
\includegraphics[width=150mm]{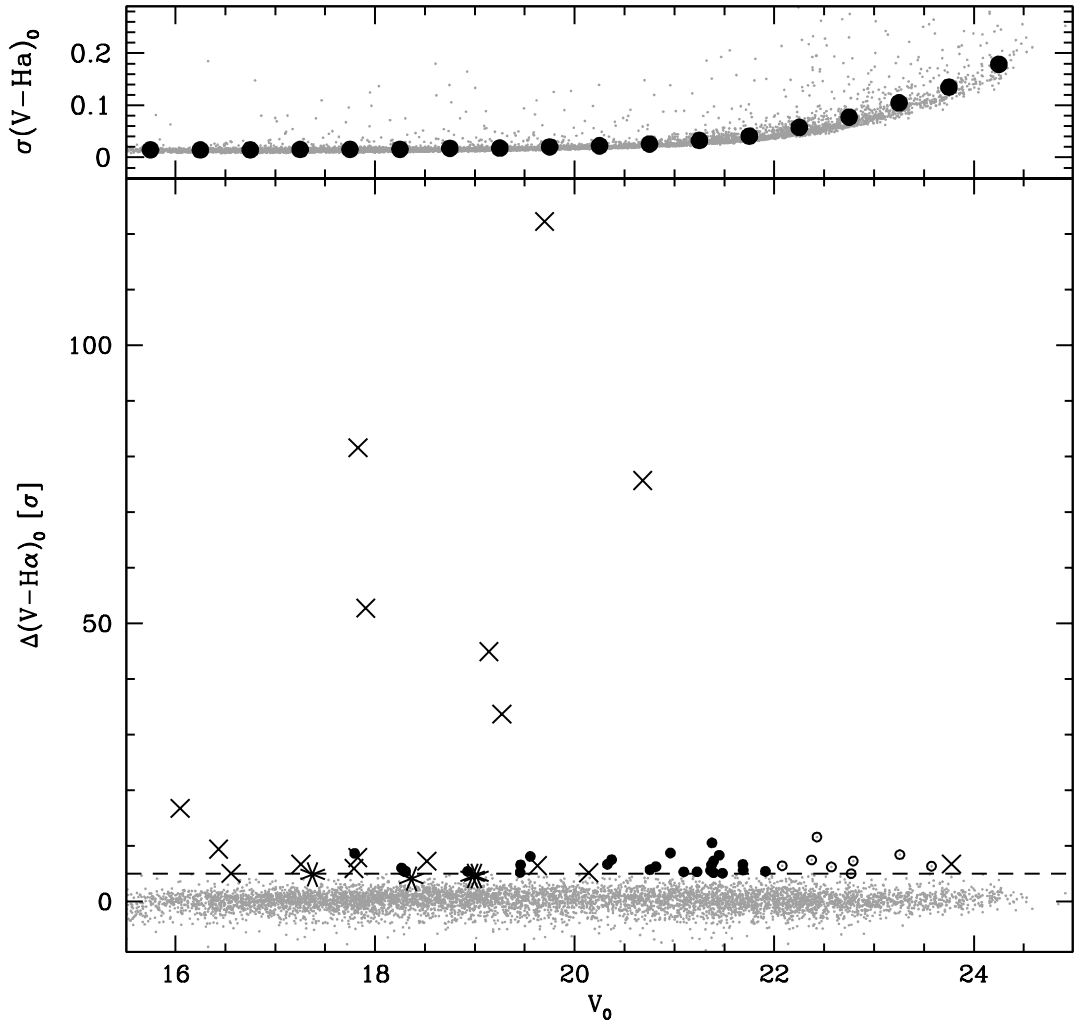}
\caption{{\it Upper panel}: distribution of photometric errors as function of the magnitude $V_0$.  
The large filled circles represent the average value of the photometric error of each 0.5 magnitude-wide bin. 
{\it Lower panel: } Color excess measured for each single star in unit of $\sigma$ (upper panel). 
The dashed line corresponds to the 5 sigma threshold above which we selected the Halpha candidate emitters. 
The black crosses and dots correspond to the  objects with and without the X-ray counterpart, respectively. 
The faintest objects ($V_0>22$), for which the classification is uncertain because of the large photometric errors, are plotted as open circles.  
The four objects marked by an asterisk correspond to X-ray counterparts selected  (see Section \ref{sec:photo} for more details).}
\label{selBW}
\end{figure*}

We used the method described in \citet{demarchi10} and recently applied to GCs by \citet{beccari13} and \citet{pallancaM28I} in order to identify all objects showing $H\alpha$ excess in the WFPC2 field of view. 
In Figure \ref{colcol} we show as gray points the position of all the stars with magnitude fainter than the MS Turn off in the $(V-H\alpha)_0$ vs $(V-I)_0$ color-color diagram. 
Note that with this color combination we can both reproduce the continuum for different spectral types and provide a good estimate of the $H\alpha$ emission, being the $H\alpha$ line contribution to the $V$ band negligible. 
As expected, the vast majority of MS stars  shows very low (if any) $H\alpha$ excess emission. 
As a consequence in the $(V-H\alpha)_0$ vs $(V-I)_0$ plane they define a very narrow sequence, which empirically indicates the locus of stars with no $H\alpha$ emission. 
The conversion of the $(V-I)_0$ colors into stellar temperatures and spectral types indicated in the figure is done using the atmospheric model of~\citet[][gray triangles]{bessell98}. 
To first select objects with $H\alpha$ excess and then estimate the pEW of the emission  we calculated a reference line, for the stars not showing  $H\alpha$ excess, defined as the median $(V-H\alpha)_0$ color of stars with a combined photometric error (computed as $\sigma_ {(V-H\alpha)_0} = \sqrt{\sigma_V^2 +\sigma_{H\alpha}^2}$) smaller than 0.05 mags. 
It is important to note that the empirical relation (gray solid line in the Figure \ref{colcol}) agrees very well with the theoretical one from the atmospheric model of \citet{bessell98}. 
For each source,  we measured the "color" difference ($\Delta (V-H\alpha)_0$) between the observed $(V-H\alpha)_0$ color and the value of $(V-H\alpha)_0$ measured along the reference line at the source $(V-I)_0$. 
We then selected as candidate $H\alpha$ emitters those objects with $\Delta (V-H\alpha)_0$  at least five times larger than the source intrinsic  error (see Figures \ref{selBW}).
The use of such a conservative threshold  guarantees to avoid the selection of objects showing a ``fake'' color excess, that is actually due to a large intrinsic photometric error. 
We performed a visual inspection of each candidate $H\alpha$ emitter onto the stacked images in order to exclude objects possibly contaminated from close saturated stars.
This procedure allowed us to reject  almost half of the initially selected candidates.

\begin{figure*}
\includegraphics[width=150mm]{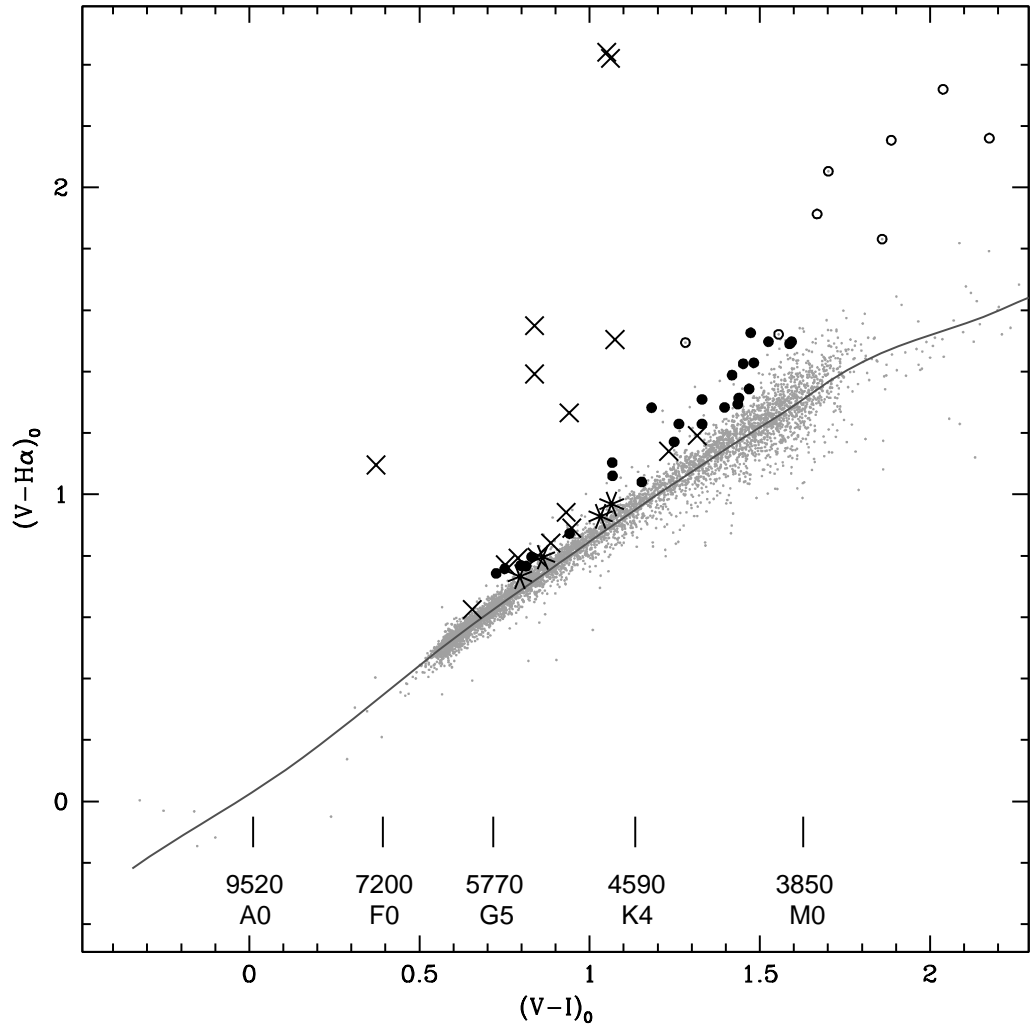}
\caption{Same as Figure \ref{colcol}. We report here the location of objects with H$\alpha$ excess. 
The black crosses (or asterisks, see Figure \ref{selBW}) and dots correspond to the  objects with and without the X-ray counterpart, respectively. 
The faintest objects ($V_0>22$), for which the classification is uncertain because of the large photometric errors, are plotted as open circles.}
\label{colcolconogg}
\end{figure*}

\begin{figure*}
\includegraphics[width=150mm]{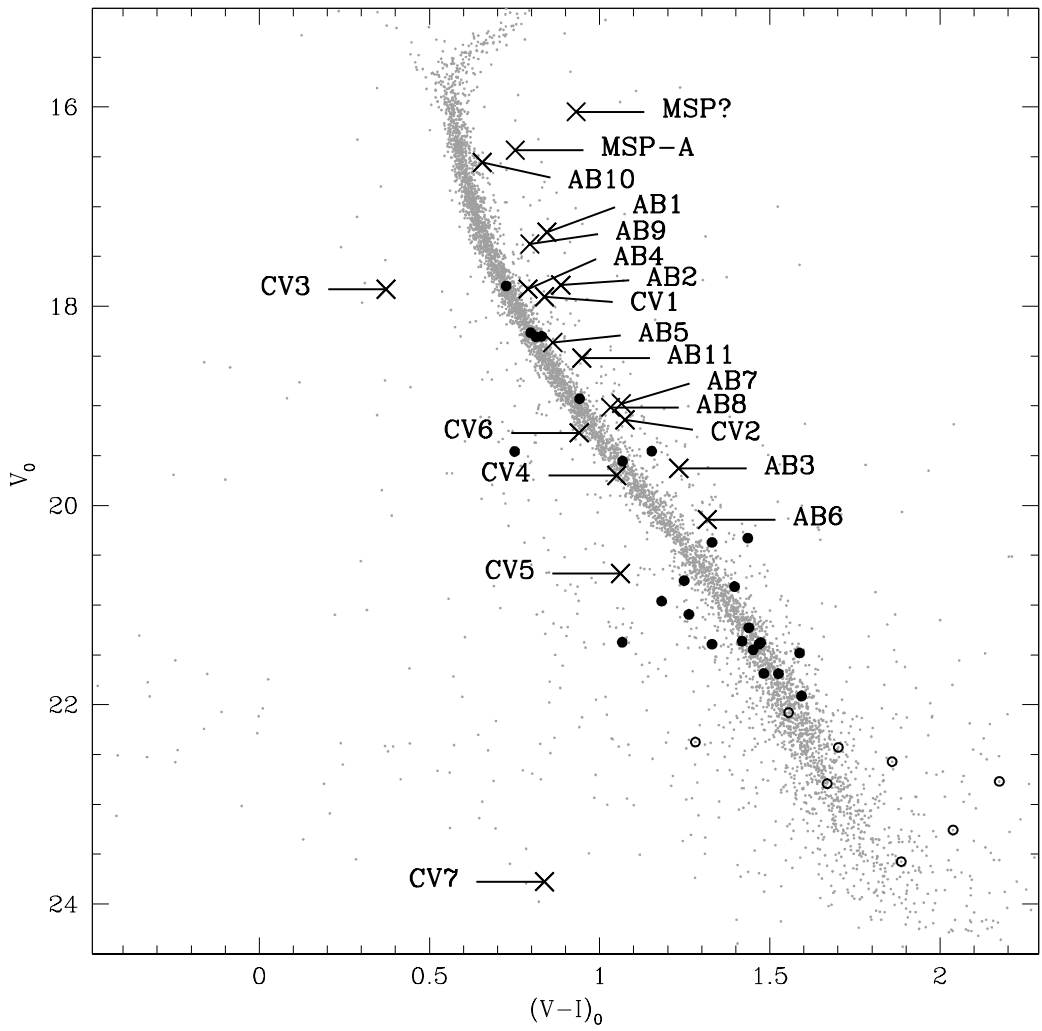}
\caption{CMD location of object with H$\alpha$ excess. 
The black crosses and dots correspond to the  objects with and without the X-ray counterpart, respectively. 
As in previous Figures, the faintest objects ($V_0>22$), for which the classification is uncertain because of the large photometric errors, are plotted as open circles.
The names of sources with X-ray counterpart are reported for clarity.}
\label{cmd}
\end{figure*}

We accepted as bona fide $H\alpha$ emitters also the objects with $(V-H\alpha)_0$ color excess 4 times larger than the intrinsic error, but falling within $0\farcs5$ from the nominal position of a X-ray source. 
In this way we added 4 more objects to the previous selection (See stars marked by asterisks in Figure \ref{selBW} and \ref{colcolconogg}). 
Following this approach we identified 53 candidate $H\alpha$ emitters: their coordinates and magnitudes are listed in Table \ref{tabella}, and their position in the $(V,V-I)$-CMD is shown in Figure \ref{cmd}. 

\begin{figure*}
\includegraphics[width=150mm]{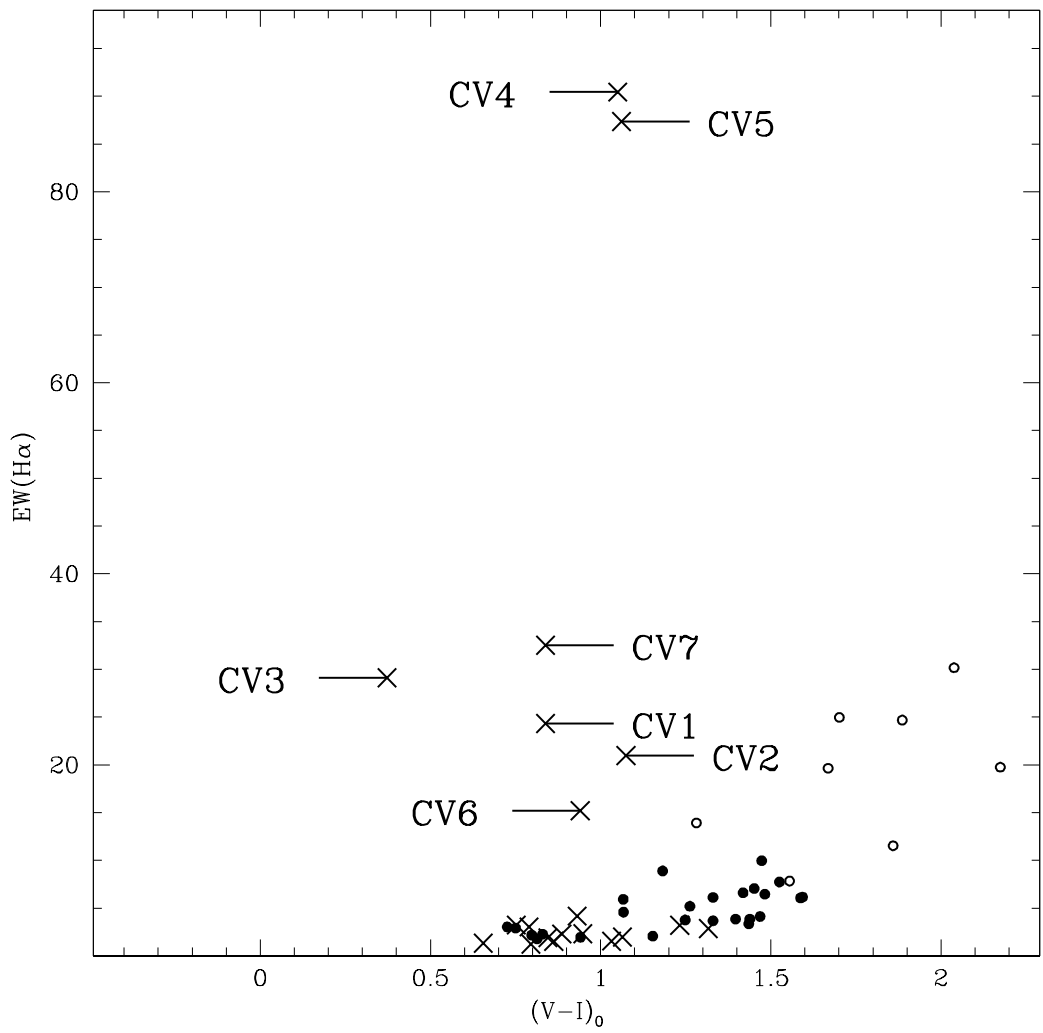}
\caption{Estimated photometric EW of the emission, as a function of the color index representative of the spectral type.
The symbols are as in the previous Figure.
Names of known CVs are reported for clarity.}
\label{ew}
\end{figure*}

We then reported the position of these objects in the color-color diagram (Figure \ref{colcolconogg}), the ideal plane to directly measure the pEW of the emission.
For all the selected  bona-fide $H\alpha$ emitters, we derived a pEW estimation of the $H\alpha$ emission line following equation (4) in \citet{demarchi10}: ${\rm pEW=RW}\times[1-10^{(-0.4\times\Delta {\rm H}\alpha)}]$,  where RW is the rectangular width  of the filter in \AA\ units, which depends on the specific characteristics of the filter \citep[see Table 4 in][]{demarchi10}. 
The obtained values are listed in Table \ref{tabella} and reported in Figure \ref{ew} as a function of the $(V-I)_0$ color.  
The 8 faintest (with $V_0>22$) objects without a X-ray counterpart, for which the photometric errors are larger than $\sigma_{(V-H\alpha)_0} >0.05$ mag are considered  "low confidence candidates" and for this reason they are plotted as open circles in Figures \ref{selBW}, \ref{colcolconogg}, \ref{cmd} and \ref{ew}. 

\subsection{Comparison with MUSE}
Very recently \citet{husser16} presented the spectroscopic analysis of more than 12,000 stars in NGC 6397 observed with the VLT-MUSE, a panoramic integral-field spectrograph able to acquire low resolution spectra ($R=2000-4000$, from the bluest to the reddest wavelengths) in the range $0.465-0.93\mu$m \citep[][]{bacon14}.  
While most of the $H\alpha$ emitters selected through our photometric strategy fall below the MUSE detection threshold, we were able to retrieve the spectra of 20 sources in common with our photometric catalog.  
They have been used to obtain a spectroscopic estimate of the $H\alpha$ EW (sEW), for a direct comparison with our photometric values (pEWs).

From the best-fit isochrone, we assigned a temperature and a surface gravity to each target. 
These have then been used to calculate a synthetic spectrum with the {\tt SYNTHE} code developed by R. L. Kurucz. 
After the convolution with a Gaussian profile, needed to reproduce the spectral resolution of MUSE, we determined the radial velocity of each target through cross correlation (by using the iraf task fxcor) with the synthetic spectrum in the Calcium triplet region, and we reported the observed spectra to the rest frame. 
In the following comparison with MS stars, the shift to the rest frame is fundamental because, being members of binary systems, the $H\alpha$ emitters can have a radial velocity significantly different from GC systemic velocity and variable along orbital phase. 
We then adopted the same approach to analyse a sample of MS stars with comparable temperature and gravity, and once reported to the zero velocity rest frame, we built a reference median template through the IRAF task scombine. 
Finally, because for most objects the $H\alpha$ line in the atmosphere of the $H\alpha$ emitters just appears as a less absorbed feature with respect to what observed in unperturbed stars with similar physical parameters, we divided the target spectra by the reference template, in order to isolate the emission component, and we measured the sEW through a Gaussian fit.
In some cases the line profile was asymmetric (likely due to $H\alpha$ emission coming from different regions of the system; e.g. see \citealp{sabbi03}) and a minor, secondary Gaussian component was used in the fit.  

In order to estimate the uncertainties in the sEW measures we performed Monte Carlo simulations. We first simulated 5 templates with different $H\alpha$  sEWs  (1, 3 ,5, 10 and 50 \AA). For each template we built a set of simulated spectra with variable signal to noise ratio (SNR), ranging  from 5 to 50 in steps of 5. Then we ran an automatic procedure able to measure the sEW in each set of simulated spectra.  Thus the dispersion of the sEW measures was derived as a function of both the measured sEW and the spectrum SNR.  This grid of values was used to derive the uncertainty of each $H\alpha$  sEW measured in a spectrum of a given SNR. The values obtained are reported in Table \ref{tabella}.

For CV1 and  eight ABs, two different MUSE spectra are available in the \citet{husser16} data-set and we thus obtained two measures. 
 For the eight ABs the two values agree each other within $1-\sigma$ uncertainties.
In the case of (the main component of) CV1, we obtain 14 and 27.1 \AA, while \citet{edmonds99} found 21 \AA\ from HST spectroscopy. 
These differences are likely due to intrinsic variability of the emission and the time difference between the observations.
All the sEWs thus obtained are listed in Table 1, where we also report the values quoted in Table 1 of \citet{edmonds99} for CVs 1--4.

\begin{figure*}
\includegraphics[width=150mm]{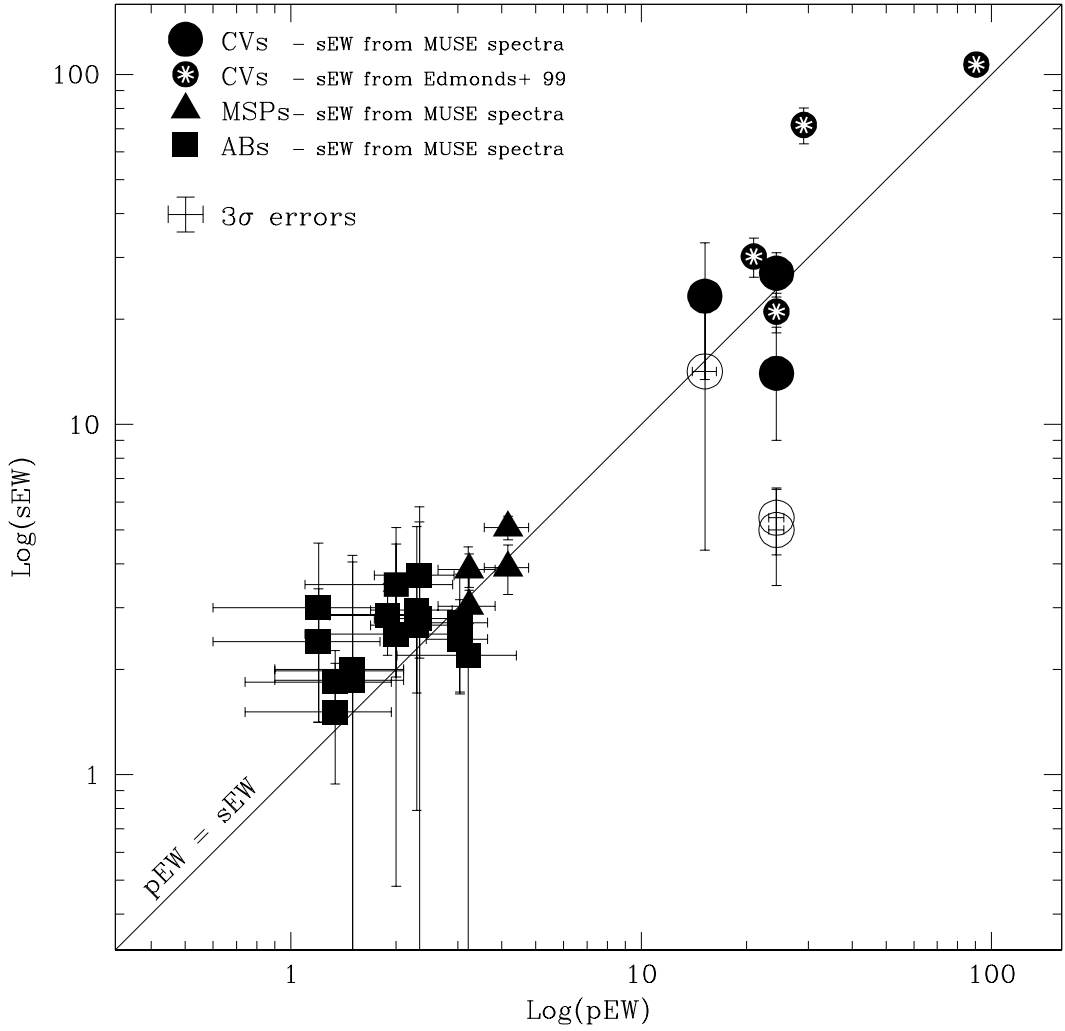}
\caption{Spectroscopic EW measured from MUSE spectra as a function of the EWs estimated from photometry. 
For most of the targets more than one spectroscopic measure is shown. 
Different symbols correspond to different classes of objects as reported in the legend.
The line corresponds to the locus of equality  pEW=sEW. A clear correlation between the two independent EW measures is present in particular in the AB/MSP region. 
In the part of the plot were CVs are located the scatter is larger likely because of variations of the EW along the time.
Note that in the Figure we plotted 3-$\sigma$  error bars, while the uncertainties reported in Table \ref{tabella} are 1-$\sigma$ errors.
}
\label{correw}
\end{figure*}

As shown in Figure \ref{correw}, we find a general good agreement between the photometric and the spectroscopic measurements of the $H\alpha$ EW for the stars in common with the MUSE catalog. 
The largest discrepancies are found for CVs (note that the axes in the figure are logarithmic), and can again be explained by the intrinsic variability of their $H\alpha$ emission and by taking into account that the WFPC2 and the MUSE observations of the cluster have been performed in very different epochs (the photometric and the spectroscopic data  were acquired in April 1999 and July 2014, respectively).  
This comparison is critical since it provides the first direct test of the fact that a combination of broad band with narrow band $H\alpha$ imaging allows to obtain reliable photometric measurements of EW of stars showing $H\alpha$ excess. This is particularly critical in regions where  severe stellar crowding or target faintness prevents to obtain reliable spectroscopic data.  

\section{Discussion}
Using the selection criteria described in the previous section we found 53 objects with $H\alpha$ excess. 
Through  cross-correlation with catalogs of peculiar sources in this GC (i.e. X-ray sources, and variable stars; \citealp{cohn10} and \citealp{kaluzny03}, respectively) we were able to find associations with previously known objects.
In particular we found that 20 $H\alpha$ emitters  (7 CVs, 2 MSPs and 11 ABs) have an X-ray counterpart. 
Moreover, 8 objects (among the brightest) were identified as variable stars and  9 sources were previously classified  as BY Dra-type binaries. 
The remaining 33 candidates are new sources. We stress here that  the eight faintest objects (without a X-ray counterpart) are however considered uncertain because of the large photometric errors.
See Table \ref{tabella} for a summary.

All the 7 detected CVs show a pEW larger than 15 \AA. 
In particular they appear to be divided in two groups: CV4 and CV5  have pEW larger than 80 \AA, while CV1, CV2, CV3, CV6 and CV7 show an average $H\alpha$ excess (pEW$\sim 25$\AA). 
Such a dichotomy is  suggestive  of a different evolutionary state \citep{pallancaM28I} or a different accretion state (likely depending on the  nature of the compact object). 
In fact, even if they are classified as CVs, we cannot definitely rule out the hypothesis  that some of them could be T-MSPs, similar to  PSR J1023+0038 and PSR J1227$-$4853 \citep{stappers14,roy15}.
In the CMD  (see Figure \ref{cmd}) they are located both on the left and on the right side of the MS. 
This evidence confirms that the CMD position alone (e.g. a color  bluer than the MS) cannot be exhaustively used to select  a complete sample of CVs and that the pEW that we can measure with the observational method described in this paper is indeed critical.

Interestingly all the 11 ABs found in our catalog show a pEW smaller than 4 \AA\ and lie on the sequence of  binary systems observed on the red side of the MS (see Figure \ref{cmd}). 
Note also that while they all have X-ray counterparts  known,  only 7 were previously identified in optical bands as  possible BY Dra binaries (\citealp{taylor01}; see also Table \ref{tabella}), thus suggesting that the $H\alpha$ selection method allowed us to detect 4 more ABs ($\sim 50\%$ more) with respect to the optical variability study.

\begin{figure*}
\includegraphics[width=150mm]{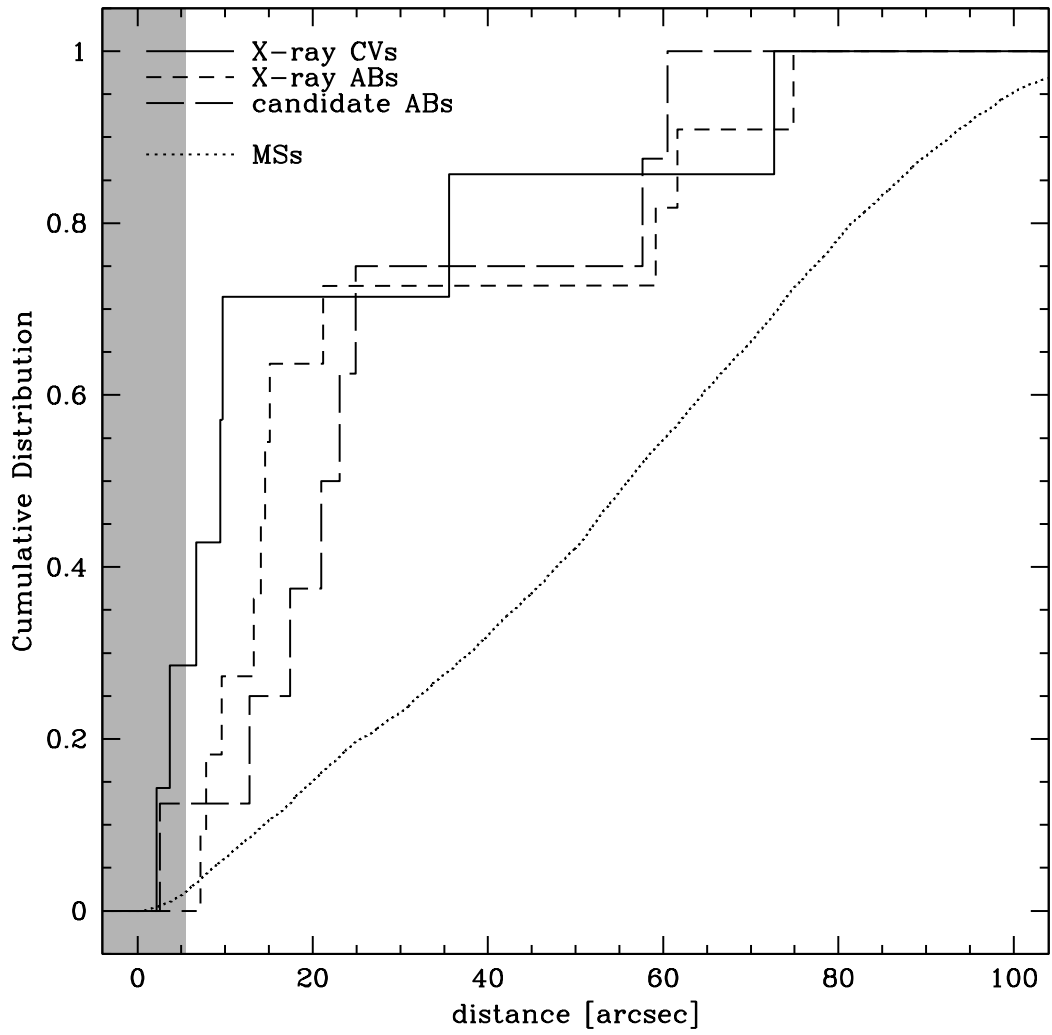}
\caption{Cumulative radial distribution of different populations. 
In the top-left part of the plot the legend of different populations  is reported.
The shaded gray area represents the core radius \citep[5.5\arcsec ;][]{cohn10}. 
As evident the X-ray CVs and ABs as well as the candidate ABs are significantly more concentrated than MSs. }
\label{cumulativa}
\end{figure*}

We performed a Kolmogorov-Smirnov test comparing the radial distributions of X-ray CVs, X-ray ABs and MS stars. 
We found that both the CVs and the ABs are more concentrated than MS stars (see Figure \ref{cumulativa}). 
In particular the probability that they are extracted from the same parent population  of MSs is only  of the order of 0.1$-$0.2\%, while the  radial distributions of ABs is compatible with that of CVs with a probability of 28\%.  
Finally, the two MSPs (the radio known U12 and the candidate U18) have a small pEW, slightly larger than the values measured for the ABs.
 This is a further confirmation that source U18 is very similar to U12 (PSR J1740$-$5340A) and it well fits within the scenario already proposed by \citet{bogdanov10} who  found that its X-ray and optical properties are consistent with those of a binary containing a rotation-powered pulsar wind interacting with material from the secondary star.

Among the sources without a X-ray counterpart, only one object is known in the literature and it is classified as a possible BY Dra-type binary (star number 10 in Table \ref{tabella}). 
Eleven of the H$\alpha$ emitters with no X-ray counterpart have a pEW$<4$\AA\ and a CMD position compatible with the sequence of binaries. 
This suggests that they are likely ABs. 
The remaining objects, showing pEW  larger than  $4$\AA\  could be unknown  interacting binaries (MSPs or CVs). 
Their classification remains uncertain at the moment, however they surely are  primary targets for further investigation.

\section{Summary and future perspectives}
By using  a set of high-resolution images of the central region of the GC NGC 6397, acquired through the $V$, $I$, $H\alpha$ filters we identified 53 objects with $H\alpha$ excess. 
Among them there are 20 X-ray sources: 7 CVs, 11 ABs, 1 radio MSP and 1 candidate MSP.
In particular all the ABs have a pEW smaller than 4 \AA\ and in the CMD  they are located onto the cluster binary sequence. 
The two MSPs are both located below the Sub Giant Branch and have a pEW of emission slightly larger than that of ABs. 
The CVs are mostly out of canonical sequences in the CMD, both on the left and on the right side of the MS, and they have a significantly large $H\alpha$ emission (pEW>15\AA) suggesting that for these objects the source of emission is an accretion process, instead of cromospheric activity (as likely is in ABs). 
The CVs seem to have two different regimes of accretion. 
In particular CV4 and CV5 are characterised by a strong $H\alpha$ similarly to the peculiar T-MSP J1824$-$2452I in M28 \citep{pallancaM28I}.
We also detected several objects with no X-ray counterparts: these objects can be ABs and or interacting binaries.

In this paper we used an innovative method to estimate the EW through photometric observations. 
The derived values have been found to nicely correlate with the spectroscopic measurements, thus confirming the reliability of the photometric approach. 
Such a method  allows to detect new objects with $H\alpha$ excess and, in principle, to measure the emission EW also from objects which are too faint to be target of spectroscopic observations.

Note that the proposed selection method could play a fundamental role in the detection of T-MSPs. 
For example the two known field T-MSPs were initially misidentified as magnetic CVs \citep{bond02,butters08}. 
In particular T-MSPs during their life likely changes significantly the amount of H$\alpha$ excess (i.e. during quiescent state PSR J1824$-$2452I shows a pEW of a few \AA, as the ABs, while during burst phase it is characterised by strong emission, even larger than some  CVs).

An important step forward will be to obtain similar dataset in different epochs in order to identify objects (as T-MSPs, active LMXB and CVs)  that are significantly changing status along time.
Moreover also coordinated photometric and spectroscopic observations planned at the same time would be important in order to further test the correlation between pEW and sEW without the bias of intrinsic variability of accretion.

\section*{Acknowledgments}
We thank the anonymous referee for the careful reading of the manuscript and useful comments.
C.P. warmly thanks the support of the ESO Visitor Program.
The research leading to these results has received funding from the European Community's Seventh Framework Programme (/FP7/2007-2013/) under grant agreement No 229517.  
Based on observations made with the NASA/ESA Hubble Space Telescope, obtained from the data archive at the Space Telescope Institute. 
STScI is operated by the association of Universities for Research in Astronomy, Inc. under the NASA contract NAS 5-26555.  



{\it Facilities:} \facility{ESO (WFI)}, \facility{HST (WFPC2)}.



\newpage
\begin{table}
\begin{center}
\begin{tabular}{|l|cc|ccc|c|c|ccclr|}
\hline
ID        & RA                 & Dec             &$V_0$&$I_0$&H$\alpha_0$& pEW                & sEW          & B10 & G01 & K06    & T01          &   H16     \\
       & J2000             & J2000         &           &         &                    & \AA                  & \AA         &         &        &            &                  &             \\
\hline
\hline
CV1     & 265.1733338 & -53.6720091 &  17.91 &  17.07 &  16.51 &  24.3$\pm$0.4  &  14.0$\pm$1.7-5.0$\pm$0.5      &  U23  & CV1 & V12  &                       & 10639  \\   
            &                       &                      &            &            &            &                          &  27.1$\pm$1.3-5.4$\pm$0.4     &           &         &         &                       &             \\   
            &                       &                      &            &            &            &                          &  21$^E$$\pm$   0.9            &           &         &         &                       &             \\   
CV2     & 265.1762738 & -53.6746139 &  19.14 &  18.07 &  17.64 &  21.0$\pm$0.4  &   30.2$^E$$\pm$  1.3        &   U19 & CV2 & V34  &                       &             \\   
CV3     & 265.1777106 & -53.6720105 &  17.83 &  17.46 &  16.73 &  29.1$\pm$0.3  &  71.7$^E$$\pm$  2.8          &   U17 & CV3 & V33  &                       &             \\   
CV4     & 265.1742981 & -53.6725857 &  19.70 &  18.65 &  17.26 &  90.4$\pm$1.0  &  106.7$^E$$\pm$  1.7            &   U21 & CV4 &         &                       &              \\  
CV5     & 265.1737722 & -53.6747031 &  20.68 &  19.62 &  18.26 &  87.4$\pm$1.4  &                                           &   U22 & CV5 &         &                       &              \\  
CV6     & 265.2040901 & -53.6634800 &  19.27 &  18.33 &  18.01 &  15.2$\pm$0.4  &  14.2$\pm$3.3-23.3$\pm$3.3    &  U10 & CV6 & V13 &                       &   8919   \\  
CV7     & 265.1907295 & -53.6782007 &  23.78 &  22.94 &  22.23 &  32.5$\pm$5.5  &                                          &   U11 & CV7 &         &                       &              \\  
\hline
MSP-A & 265.1859047 & -53.6782108 &  16.43 &  15.68 &  15.66 &    3.2$\pm$0.2  &  3.0$\pm$0.1                  &  U12  &         & V16  & BY (WF4-1)  &   4991   \\  
            &                       &                      &            &             &            &                          &  3.8$\pm$ 0.2                 &         &         &         &                      &               \\ 

MSP?  & 265.1775326 & -53.6743174 &  16.05 &  15.12 &  15.11  &    4.2$\pm$0.2  &  3.9$\pm$ 0.2             &  U18 &          & V31 & BY  (PC-8)  ?&   5596    \\ 
            &                       &                      &            &            &             &                          &  5.1$\pm$ 0.1            &          &         &         &                      &               \\ 
\hline
AB1     & 265.1788187 & -53.6760517 &  17.26 &  16.41 &  16.46  &    1.9$\pm$0.2  &  2.9$\pm$ 0.2             & U15 &         &         & BY (PC-2)     &   5489    \\ 
            &                       &                      &            &            &             &                          &  2.9$\pm$  0.2            &          &         &         &                      &              \\ 
AB2     & 265.1689240 & -53.6730075 &  17.79 &  16.90 &  16.95  &    2.3$\pm$0.2  &  2.9$\pm$  0.3            &  U43 &         &         & BY (PC-4)     & 10919   \\  
            &                       &                      &            &            &             &                          &  3.0$\pm$  0.7           &          &         &         &                      &              \\ 
AB3     & 265.1669239 & -53.6712405 &  19.63 &  18.40 &  18.49  &    3.2$\pm$0.4  &  2.2$\pm$  0.8            & U67 &         &         &                      & 11027    \\ 
AB4     & 265.1702716 & -53.6714234 &  17.83 &  17.04 &  17.04  &    3.0$\pm$0.2  &  2.4$\pm$  0.2            &  U69 &         &         & BY (PC-5)     & 10841   \\  
            &                       &                      &            &            &             &                          &  2.7$\pm$  0.3            &          &         &         &                      &             \\ 
AB5*    & 265.1779063 & -53.6579716 &  18.36 &  17.50 &  17.57  &    1.5$\pm$0.2  &  2.0$\pm$ 0.4             & U73 &         &         &                      & 14084   \\  
AB6     & 265.1818423 & -53.6751222 &  20.14 &  18.83 &  18.95  &    2.9$\pm$0.5  &                                      &  U75 &         &         &                      &              \\ 
AB7*    & 265.1936621 & -53.6875928 &  18.98 &  17.91 &  18.01  &    2.0$\pm$0.3  &  2.5$\pm$  0.7           &  U81  &         & V14 &                      &   4580   \\  
            &                       &                      &            &            &             &                          &  3.5$\pm$  0.5            &         &         &         &                      &             \\ 
AB8*    & 265.2022836 & -53.6609875 &  19.02 &  17.98 &  18.09  &    1.5$\pm$0.2  &  1.9$\pm$  0.8            & U82 &         &         & BY (WF2-1)  &   8990   \\ 
            &                       &                      &            &            &             &                          &  2.0$\pm$  0.7            &         &         &         &                      &              \\ 
AB9*    & 265.1786817 & -53.6739516 &  17.37 &  16.58 &  16.64  &    1.2$\pm$0.2  &   2.4$\pm$ 0.3            &  U87 &         &         & BY (PC-3)     &   5499  \\ 
            &                       &                      &            &            &             &                          &  3.0$\pm$  0.5            &         &         &         &                      &             \\ 
AB10   & 265.1784915 & -53.6731722 &  16.56 &  15.90 &  15.93  &    1.3$\pm$0.2  &   1.5$\pm$  0.2           &  U88 &         & V19  & BY (PC-1)     & 10230 \\ 
            &                       &                      &            &            &             &                          &  1.8$\pm$   0.1           &          &         &         &                      &            \\ 
AB11   & 265.1741289 & -53.6707139 &  18.52 &  17.57 &  17.63  &    2.3$\pm$0.2  &   2.8$\pm$  1.0           &  U90 &         &         & BY (PC-6)     & 10564 \\ 
            &                       &                      &            &            &             &                          &  3.7$\pm$   0.5           &          &         &        &                       &            \\ 
\hline
\ \ 1    & 265.1826286 & -53.6782726 &  17.80 &  17.07 &  17.05 &   3.0$\pm$0.2  &   N.M.   & & & &                         &     5213  \\ 
\ \ 2    & 265.1792970 & -53.6770140 &  18.27 &  17.47 &  17.50 &   2.2$\pm$0.3  &   N.M.   & & &  &                        &   5438   \\  
\ \ 3    & 265.1669335 & -53.6582910 &  18.30 &  17.47 &  17.51 &   2.3$\pm$0.3  &   N.M.   & &  & &                        &  14481 \\  
\ \ 4    & 265.1869772 & -53.6735069 &  18.31 &  17.49 &  17.54 &   1.8$\pm$0.2  &   N.M.   &  & & &                        &   9678   \\  
\ \ 5    & 265.1721989 & -53.6682152 &  18.93 &  17.99 &  18.06 &   2.0$\pm$0.3  &   N.M.    & & & &                        &  10681 \\  
\ \ 6    & 265.1743181 & -53.6746601 &  19.45 &  18.30 &  18.41 &   2.1$\pm$0.3  &   N.M.   & & & &                         &    5917 \\  
\ \ 7    & 265.2040749 & -53.6776261 &  19.46 &  18.71 &  18.70 &   2.9$\pm$0.3  &              & & & &                         &             \\  
\ \ 8    & 265.1684970 & -53.6574071 &  19.56 &  18.49 &  18.50 &   4.6$\pm$0.5  &              & & & &                         &             \\  
\ \ 9    & 265.1835070 & -53.6748396 &  20.33 &  18.89 &  19.03 &   3.4$\pm$0.4  &              & & & &                         &             \\  
10   & 265.1982481 & -53.6658034 &  20.37 &  19.04 &  19.14 &   3.7$\pm$0.4  &              & & & &   BY (WF3-4)   &             \\  
11   & 265.2118907 & -53.6617568 &  20.75 &  19.51 &  19.58 &   3.8$\pm$0.5  &               & & & &                        &              \\  
12   & 265.1803621 & -53.6782471 &  20.81 &  19.42 &  19.53 &   3.9$\pm$0.5  &              & & & &                         &              \\  
13   & 265.1763986 & -53.6829417 &  20.96 &  19.78 &  19.68 &   8.9$\pm$0.9  &              & & & &                         &               \\ 
14   & 265.2131393 & -53.6744206 &  21.09 &  19.83 &  19.86 &   5.2$\pm$0.8  &              & & & &                         &             \\  
15   & 265.2074017 & -53.6694465 &  21.23 &  19.79 &  19.91 &   3.9$\pm$0.6  &              & & & &                         &             \\  
16   & 265.1715432 & -53.6897092 &  21.36 &  19.94 &  19.97 &   6.6$\pm$0.9  &              & & & &                         &               \\ 
17   & 265.1748629 & -53.6915474 &  21.37 &  20.31 &  20.27 &   5.9$\pm$0.8  &              & & & &                         &               \\ 
18   & 265.1691896 & -53.6571696 &  21.38 &  19.90 &  19.85 &  10.0$\pm$1.0 &              & & & &                         &             \\  
19   & 265.1848957 & -53.6763541 &  21.39 &  19.92 &  20.05 &   4.1$\pm$0.7  &              & & & &                         &              \\  
20   & 265.2063740 & -53.6654454 &  21.39 &  20.06 &  20.08 &   6.1$\pm$0.7  &              & & & &                         &             \\  
21   & 265.2089910 & -53.6693385 &  21.45 &  20.00 &  20.02 &   7.1$\pm$0.7  &              & & & &                         &             \\  
22   & 265.1824425 & -53.6817866 &  21.48 &  19.89 &  19.99 &   6.1$\pm$0.9  &              & & & &                         &               \\ 
23   & 265.1929464 & -53.6694839 &  21.69 &  20.16 &  20.19 &   7.7$\pm$1.1  &              & & & &                         &             \\  
24   & 265.2055337 & -53.6651658 &  21.69 &  20.20 &  20.26 &   6.5$\pm$0.9  &              & & & &                         &             \\  
25   & 265.1990390 & -53.6602075 &  21.91 &  20.32 &  20.41 &   6.1$\pm$1.1  &              & & & &                         &             \\  
26 ?  & 265.2121746 & -53.6743002 &  22.08 &  20.52 &  20.56 &   7.8$\pm$1.1  &              & &   & &                        &             \\  
27 ?  & 265.1810287 & -53.6956426 &  22.37 &  21.09 &  20.88 &  13.9$\pm$1.8 &              & &  & &                        &               \\ 
28 ?   & 265.1688267 & -53.6575723 &  22.43 &  20.73 &  20.38 &  25.0$\pm$2.6 &              & &  & &                       &             \\  
29 ? & 265.1888864 & -53.6906782 &  22.57 &  20.71 &  20.74 &  11.5$\pm$1.6 &              & &  & &                      &               \\ 
30 ?  & 265.1739928 & -53.6769901 &  22.77 &  20.60 &  20.61 &  19.8$\pm$2.7 &              & &  & &                       &             \\  
31 ?   & 265.1918767 & -53.6491195 &  22.79 &  21.12 &  20.88 &  19.6$\pm$2.4 &              & &  & &                       &             \\  
32 ? & 265.1921196 & -53.6776013 &  23.26 &  21.22 &  20.94 &  30.2$\pm$3.1 &              & &   & &                      &               \\ 
33 ? & 265.1872843 & -53.6910091 &  23.58 &  21.69 &  21.42 &  24.7$\pm$3.6 &              & &  & &                      &               \\ 
\hline
\end{tabular}
\end{center}
\caption{}{List of selected objects with H$\alpha$ excess. 
From left to right: Source Name, absolute coordinates, de-reddened magnitudes, photometric EW, spectroscopic EW (more than one value is reported if the line profile has been fitted by two gaussians, a second raw is added when more that one spectrum is available, the values marked by an ``E" are the measures from \citealt{edmonds99}, N.M. means that although the spectra are available, has been impossible to measure the EW), list of IDs in papers available in the literature: B10=\citealp{bogdanov10}, G01=\citealp{grindlay01}, K06=\citealp{kaluzny06} T01=\citealp{taylor01} H16 \citealp{husser16}. In first section of the table we report all the objects having a X-ray counterpart divided in CVs MSPs and ABs. 
The four objects marked by the * correspond to those stars selected although they were slightly below the imposed threshold (see Section \ref{sec:photo} for more details). 
The eight objects marked by the ? correspond to the faintest stars for which the classification is uncertain due to their large photometric errors.}
\label{tabella}
\end{table}

\newpage

\newpage

\newpage

\newpage

\newpage

\newpage

\newpage

\newpage


\begin{thebibliography}{}

\bibitem[Alpar et al.(1982)]{alpar82} Alpar, M.~A., Cheng, A.~F., Ruderman, M.~A., \& Shaham, J.\ 1982, \nat, 300, 728 

\bibitem[Bailyn(1995)]{bailyn95} Bailyn, C.~D.\ 1995, \araa, 33, 133

\bibitem[Bhattacharya \& van den Heuvel(1991)]{bhattvan91} Bhattacharya, D., \& van den Heuvel, E.~P.~J.\ 1991, \physrep, 203, 1 

\bibitem[Bacon et al.(2014)]{bacon14} Bacon, R., Vernet, J., Borisova, E., et al.\ 2014, The Messenger, 157, 13 

\bibitem[Beccari et al.(2006)]{beccari06} Beccari, G., Ferraro, F.~R., Possenti, A., et al.\ 2006, \aj, 131, 2551 

\bibitem[Beccari et al.(2014)]{beccari13} Beccari, G., De Marchi, G., Panagia, N., \& Pasquini, L.\ 2014, \mnras, 437, 2621 

\bibitem[Bessell et al.(1998)]{bessell98} Bessell, M.~S., Castelli, F.,   \& Plez, B.\ 1998, \aap, 333, 231

\bibitem[Bogdanov et al.(2010)]{bogdanov10} Bogdanov, S., van den Berg, M., Heinke, C.~O., et al.\ 2010, \apj, 709, 241

\bibitem[Bond et al.(2002)]{bond02} Bond, H.~E., White, R.~L., Becker, R.~H., \& O'Brien, M.~S.\ 2002, \pasp, 114, 1359 

\bibitem[Butters et al.(2008)]{butters08} Butters, O.~W., Norton, A.~J., Hakala, P., Mukai, K., \& Barlow, E.~J.\ 2008, \aap, 487, 271 

\bibitem[Cadelano et al.(2015)]{cadelanoM71A} Cadelano, M., Pallanca, C., Ferraro, F.~R., et al.\ 2015, \apj, 807, 91 

\bibitem[Cohn et al.(2010)]{cohn10} Cohn, H.~N., Lugger, P.~M., Couch, S.~M., et al.\ 2010, \apj, 722, 20

\bibitem[Cool et al.(1995)]{cool95} Cool, A.~M., Grindlay, J.~E., Cohn, H.~N., Lugger, P.~M., \& Slavin, S.~D.\ 1995, \apj, 439, 695

\bibitem[Cool et al.(1998)]{cool98} Cool, A.~M., Grindlay, J.~E., Cohn, H.~N., Lugger, P.~M., \& Bailyn, C.~D.\ 1998, \apjl, 508, L75

\bibitem[De Marchi et al.(2010)]{demarchi10} De Marchi, G., Panagia,   N., \& Romaniello, M.\ 2010, \apj, 715, 1

\bibitem[Davies(2005)]{davies05} Davies, M.\ 2005, Advances in astronomy: From the Big Bang to the Solar System, 1, 245 

\bibitem[Dotter et al.(2008)]{dotter08} Dotter, A., Chaboyer, B., Jevremovi{\'c}, D., et al.\ 2008, \apjs, 178, 89-101 

\bibitem[Edmonds et al.(1999)]{edmonds99} Edmonds, P.~D., Grindlay, J.~E., Cool, A., et al.\ 1999, \apj, 516, 250

\bibitem[Ferraro et al.(2001)]{ferraro01com6397} Ferraro, F.~R., Possenti, A., D'Amico, N., \& Sabbi, E.\ 2001, \apjl, 561, L93 
 
\bibitem[Ferraro et al.(2003)]{ferraro03} Ferraro, F.~R., Possenti, A., Sabbi, E., et al.\ 2003, \apj, 595, 179 

\bibitem[Ferraro et al.(2009)]{ferraro09} Ferraro, F.~R., Beccari, G., Dalessandro, E., et al.\ 2009, \nat, 462, 1028

\bibitem[Ferraro et al. (2012)]{ferraro12} Ferraro, F. R., et al.,  2012, \nat, 492, 393

\bibitem[Ferraro et al.(2015a)]{ferraro15a} Ferraro, F.~R., Lanzoni, B., Dalessandro, E., Mucciarelli, A., \& Lovisi, L.\ 2015a, Ecology of Blue Straggler Stars, 99

\bibitem[Ferraro et al.(2015b)]{ferraro15b} Ferraro, F.~R., Pallanca, C., Lanzoni, B., et al.\ 2015b, \apjl, 807, L1 

\bibitem[Gratton et al.(2003)]{gratton03} Gratton, R.~G., Bragaglia, A., Carretta, E., et al.\ 2003, \aap, 408, 529 
  
\bibitem[Grindlay et al.(1995)]{grindlay95} Grindlay, J.~E., Cool, A.~M., Callanan, P.~J., et al.\ 1995, \apjl, 455, L47

\bibitem[Grindlay(1999)]{grindlay99} Grindlay, J.~E.\ 1999, Annapolis Workshop on Magnetic Cataclysmic Variables, 157, 377

\bibitem[Grindlay et al.(2001)]{grindlay01} Grindlay, J.~E., Heinke, C.~O., Edmonds, P.~D., Murray, S.~S., \& Cool, A.~M.\ 2001, \apjl, 563, L53

\bibitem[Harris(1996)]{harris96} Harris, W.~E.\ 1996, VizieR Online Data Catalog, 7195,  

\bibitem[Holtzman et al.(1995)]{ho95} Holtzman, J.~A., Burrows, C.~J., Casertano, S., et al.\ 1995, \pasp, 107, 1065 

\bibitem[Husser et al.(2016)]{husser16} Husser, T.-O., Kamann, S., Dreizler, S., et al.\ 2016, \aap, 588, A148 

\bibitem[Ivanova et al.(2006)]{ivanova06} Ivanova, N., Heinke, C.~O., Rasio, F.~A., et al.\ 2006, \mnras, 372, 1043

\bibitem[Kaluzny et al.(2003)]{kaluzny03} Kaluzny, J., Rucinski, S. M., \& Thompson, I. B., 2003, AJ, 125, 1546

\bibitem[Kaluzny et al.(2006)]{kaluzny06} Kaluzny, J., Thompson, I.~B., Krzeminski, W., \& Schwarzenberg-Czerny, A.\ 2006, \mnras, 365, 548 

\bibitem[Knigge(2012)]{knigge12} Knigge, C.\ 2012, \memsai, 83, 549 

\bibitem[Lanzoni et al.(2016)]{lanzoni16}  Lanzoni, B., Ferraro, F. R.,  Alessandrini, E., et al.  2016, \apjl, 833, L29
 
\bibitem[Lyne et al.(1987)]{lyne87} Lyne, A.~G., Brinklow, A.,   Middleditch, J., Kulkarni, S.~R., \& Backer, D.~C.\ 1987, \nat, 328,   399

\bibitem[Pallanca et al.(2010)]{pallancaM28H} Pallanca, C., Dalessandro, E., Ferraro, F.~R., et al.\ 2010, \apj, 725, 1165 

\bibitem[Pallanca et al.(2013)]{pallancaM28I} Pallanca, C., Dalessandro, E., Ferraro, F.~R., Lanzoni, B., \& Beccari, G.\ 2013, \apj, 773, 122 

\bibitem[Pallanca et al.(2014)]{pallancaM5C} Pallanca, C., Ransom, S.~M., Ferraro, F.~R., et al.\ 2014, \apj, 795, 29 

\bibitem[Richer et al.(2008)]{richer08} Richer, H.~B., Dotter, A., Hurley, J., et al.\ 2008, \aj, 135, 2141 

\bibitem[Roy et al.(2015)]{roy15} Roy, J., Ray, P.~S., Bhattacharyya, B., et al.\ 2015, \apjl, 800, L12 

\bibitem[Sabbi et al.(2003)]{sabbi03} Sabbi, E., Gratton, R., Ferraro, F.~R., et al.\ 2003, \apjl, 589, L41 

\bibitem[Shara et al.(2005)]{shara05} Shara, M.~M., Hinkley, S., Zurek, D.~R., Knigge, C., \& Dieball, A.\ 2005, \aj, 130, 1829
 
\bibitem[Stappers et al.(2014)]{stappers14} Stappers, B.~W., Archibald, A.~M., Hessels, J.~W.~T., et al.\ 2014, \apj, 790, 39 

\bibitem[Stetson(1987)]{st87}Stetson, P. B. 1987, \pasp, 99, 191

\bibitem[Stetson(1994)]{st94}Stetson, P. B. 1994, \pasp, 106, 250

\bibitem[Taylor et al.(2001)]{taylor01} Taylor, J.~M., Grindlay, J.~E., Edmonds, P.~D., \& Cool, A.~M.\ 2001, \apjl, 553, L169 

\bibitem[Townsley \& Bildsten(2005)]{townsley05} Townsley, D.~M., \& Bildsten, L.\ 2005, \apj, 628, 395 

\end{thebibliography}
\end{document}